# Superatomic hydrogen: achieving effective aggregation of hydrogen atoms at pressures lower than that of metallic hydrogen

Jia Fan,[1,2,*] Chenxi Wan,[1,2,*] Rui Liu,[2,*] Zhen Gong,[2] Hongbo Jing,[1,2] Baiqiang Liu,[2] Siyang Liu,[2] and Zhigang Wang[1,2,3,†]

[1]*Institute of Atomic and Molecular Physics, Jilin University, Changchun 130012, China*
[2]*Key Laboratory of Material Simulation Methods & Software of Ministry of Education, College of Physics, Jilin University, Changchun 130012, China*
[3]*Institute of Theoretical Chemistry, College of Chemistry, Jilin University, Changchun 130023, China*

*These authors contributed equally to this work.
†Corresponding author:wangzg@jlu.edu.cn (Z. W.).

Metal hydrogen exhibiting electron delocalization properties has been recognized as an important prospect for achieving controlled nuclear fusion, but the extreme pressure conditions required exceeding hundreds of GPa remain a daunting challenge. Here, we propose a model of superatomic hydrogen, aiming to reduce the pressure conditions required for the effective aggregation of elemental hydrogen atoms. High-precision ab initio calculations indicate that the pressure required to compress the $H_{13}$ system with one central atom and 12 surrounding atoms into a superatomic state is approximately two orders of magnitude lower than that of metallic hydrogen. Atomic-level analyses reveal that in the superatomic state of compressed $H_{13}$, the central H atom donates its electron, and all electrons are delocalized on the superatomic molecular orbitals, which conforms to properties of metallic hydrogen. Our discovery in principle opens up the prospect of superatomic hydrogen in areas such as nuclear fusion.

Hydrogen, the lightest element in the universe, typically exists as a gaseous diatomic molecule under ambient conditions. However, under extreme conditions especially high pressure, hydrogen atoms appear in a solid phase [1,2]. In 1935, Wigner and Huntington proposed a seminal prediction that the application of sufficient pressure could break the chemical bonds within hydrogen molecules. In this case, molecular solid hydrogen would transform into an atomic solid with delocalized free electrons, thus conferring material metallic properties on the material and constituting a theoretical cornerstone in the investigation of metallic hydrogen [3-6]. Previous studies have shown that metallic hydrogen may hold potential applications in nuclear fusion, superconductivity, and energy science, covering both fundamental and applied research [7-10]. Despite considerable efforts made over decades through high-pressure experiments, the synthesis of metallic hydrogen has not yet been recognized [1,11,12]. Even, according to theoretical predictions, the pressure required to achieve metallic hydrogen could reach about 500 GPa or more [2,13]. Such excessively difficult conditions have prompted researchers studying the superconducting properties of metallic hydrogen to explore alternative approaches. Therefore, researchers have proposed clathrate superhydrides with specific point group symmetry as a promising alternative [11,14-16]. Representative examples include the hydrogen clathrate frameworks with encapsulated metal elements, stabilized at high pressure, such as lanthanum-centered structures [17,18], high pressure synthesized $CaH_6$ clathrates [19,20], and $CeH_9$ formed at relatively low pressure by incorporating Ce into hydrogen cages [14,21]. These clathrate superhydrides can be synthesized at pressures as low as tens of GPa, which is even lower than the synthesis conditions for metallic hydrogen. Nevertheless, it is important to note that the original purpose of pursuing solid and even metallic states of elemental hydrogen was to facilitate fusion reactions. Although the superconducting properties of these structures containing central metal atoms have been enhanced, the presence of metal atoms renders these structures unsuitable for fusion applications. In any case, clathrate superhydrides provide a valuable insight: Is it possible to achieve a clathrate hydrogen framework by way of elemental hydrogen systems to reduce the difficulty of synthesizing metallic hydrogen?

An important consideration is that alkali metals, such as Li, Na and K, which belong to the same main-group as hydrogen, are capable of forming stable cluster structures. Especially, the elemental clusters of $M_{13}$ (M = Li, Na, K) are recognized as stable superatoms [22-24]. The common property of these superatoms is the delocalization of their valence electrons in superatomic molecular orbitals (SAMOs), which exhibit orbital symmetry and electronic shell structures similar to those of ordinary atoms [25-29]. Currently, superatomic research is gradually evolving from studies of isolated structures to investigations under pressure or ionized conditions [30-32]. This transition presents significant opportunities for reducing the requirements for system construction through the utilization of superatomic states.

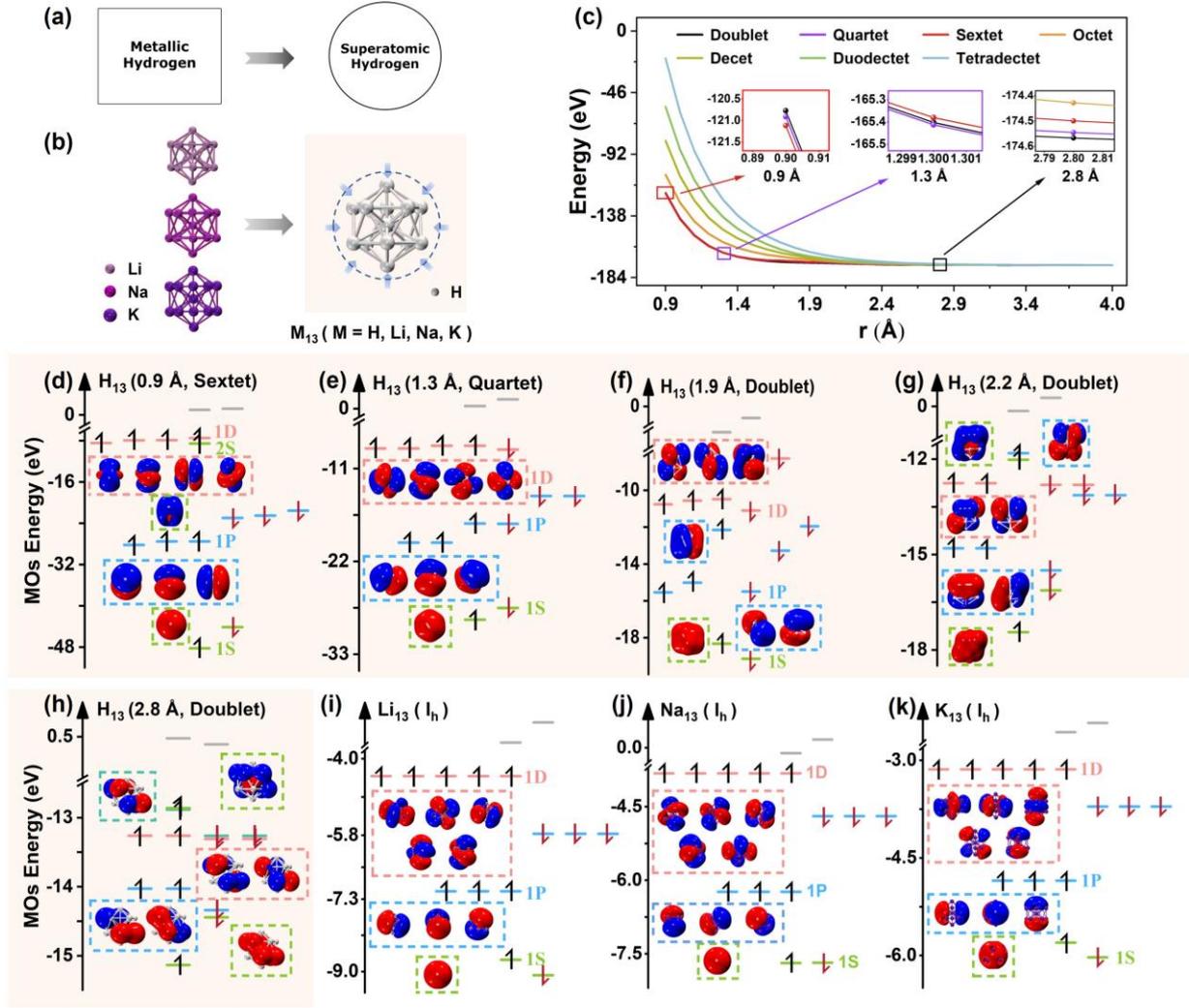

FIG. 1. Geometric and energy level structures of $M_{13}$ (M = H, Li, Na, K) systems. (a) Schematic diagram of the transformation from metallic hydrogen (left) to superatomic hydrogen (right) configurations. (b) Geometrical structures of $M_{13}$ (M = H, Li, Na, K) clusters. These structures of $H_{13}$ are obtained under confined conditions subject to isotropic pressure. (c) The compression potential energy curve of $H_{13}$ cluster, preserving $I_h$ symmetry, with the radial distance from the central hydrogen atom to the surrounding atoms varying from 4.0 Å to 0.9 Å in 0.1 Å intervals. (d)-(h) Energy levels and orbital diagrams of the $H_{13}$ clusters at radii of 0.9, 1.3, 1.9, 2.2, and 2.8 Å. (i)-(k) Energy levels and orbital diagrams of $M_{13}$ (M= Li, Na, K) clusters. These molecular orbital diagrams are given by spin-up ($\alpha$) electrons.

On the basis of above considerations, we investigate the compressive properties of a hydrogen cage composed of 13 hydrogen atoms (12 forming the outer shell and one at the center). High-precision ab initio calculations clearly reveal that this $H_{13}$ system enables an effective aggregation of bonds based on superatomic states at pressures about two orders of magnitude lower than that of traditional metallic hydrogen. Interestingly, atomic-level analyses indicate that electrons in the $H_{13}$ system are in a completely delocalized state, with the central hydrogen atom donating electrons outward, which conforms to the electronic properties of metallic hydrogen. These findings suggest that constructing hydrogen cages based on superatomic states may be a promising approach to promote nuclear fusion, thus offering a superatomic perspective for related applications such as energy and materials science.

To briefly introduce the idea of transitioning from traditional metal hydrogen with translational symmetry to superatomic hydrogen with point group symmetry (FIG. 1(a)), this study investigates the structural evolution and electronic properties of the $H_{13}$ cluster as a representative model. As shown in FIG. 1(b), referring to the $I_h$-symmetric $M_{13}$ clusters of alkali metals (M = Li, Na, K), the initial $H_{13}$ cluster was constructed with a central hydrogen atom surrounded by a shell of 12 hydrogen atoms (the detailed H—H distance evaluation was illustrated in FIG. S1 and

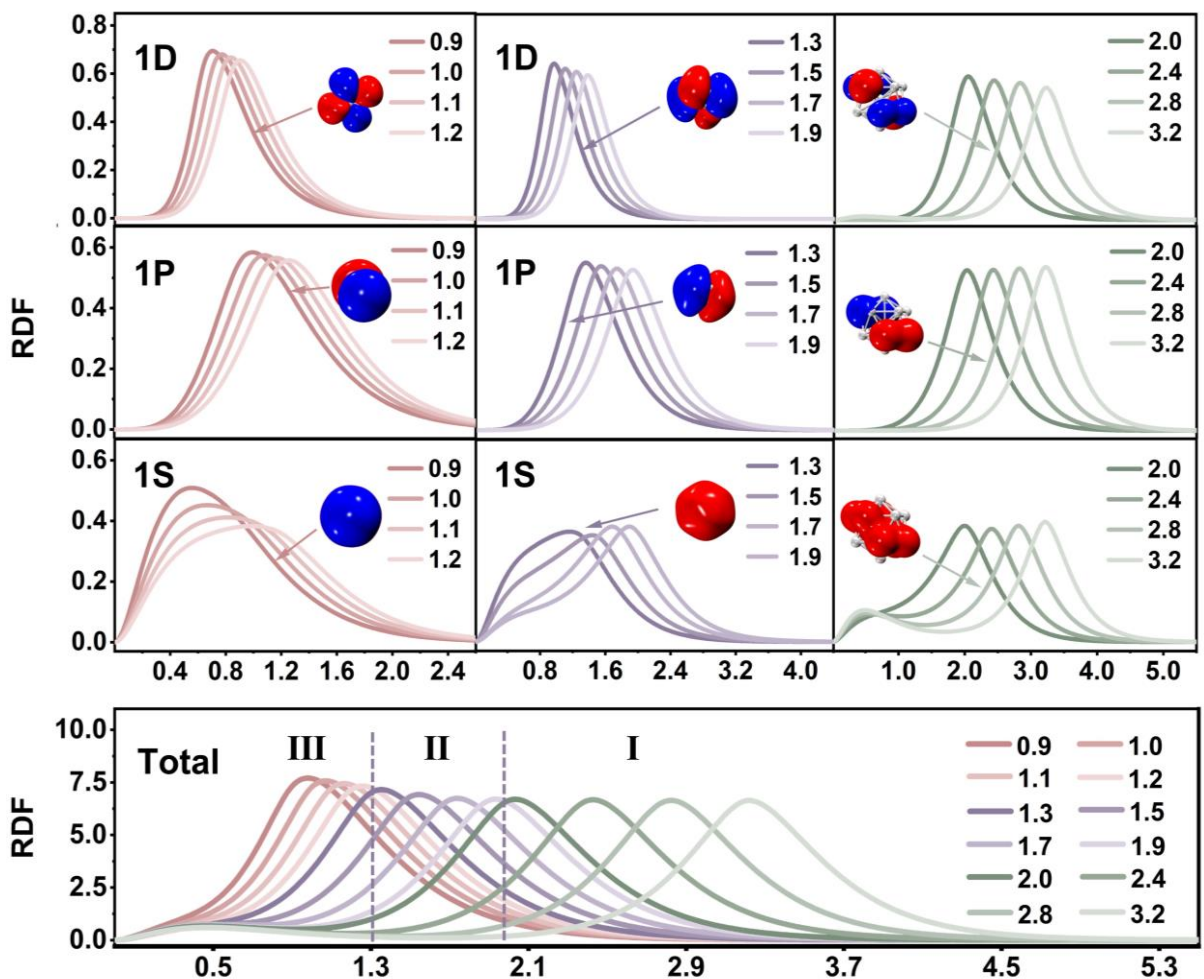

FIG. 2 Radial distribution function (RDF) of MOs in $H_{13}$ system during compression. Spherical integrals are obtained from the center of the system along the radius in RDF analysis. From top to bottom, the first three rows show the evolution of the RDF for 1D, 1P, and 1S SAMOs as the compression distance changes (in the region I, represented by the green line, these orbitals are no longer SAMOs). The bottom row displays the total RDF, calculated by spherical integration of the total electron density for each compressed configuration: $RDF = \int \rho(r) r^2 d\Omega$. Different colored lines indicate various compression. The radial coordinate ranges in orbital RDF analyses are different to accommodate structural radius variations during compression.

Table. S1 of the Supplemental Material ). Next, by progressively reducing the radial distance (r) between the central hydrogen atom and the $H_{12}$ cage from 4.0 to 0.9 Å, isotropic compression process of the $H_{13}$ system was achieved. To ensure the reliability of the results for such an electron-correlated system, the multistate complete active space second-order perturbation theory (MS–CASPT2) was used for this scan (see part 1.1 for computational details in Supplemental Material [33]). As shown in FIG. 1(c), the potential energy curves reveal a systematic evolution under different spin states. Specifically, the ground state exhibits a doublet multiplicity at r = 1.4−4.0 Å, transitions to a quartet state at r = 1.3 Å, and evolves into a sextet for r = 0.9−1.2 Å. Because the potential energy curves of these three states are very close throughout the compression process, the curves at 0.9, 1.3 and 2.8 Å were selected from different regions for local magnification to clearly depict the ground state.

Reliable electronic structures of the $H_{13}$ system at different compression regions, particularly the molecular orbital (MO) energy levels, were obtained by selecting the geometries at radial distances of 2.8, 1.9, 1.3 and 0.9 Å based on the potential energy surface scans. Single-point energy calculations for the lowest energy spin states were performed using the benchmark coupled-cluster singles and doubles with perturbative triples (CCSD(T)) method (see Supplementary part 1.2 for computational details [33]). As shown in FIG. 1(d)-(h), at radial distances of 1.3 and 0.9 Å, the MO occupation clearly exhibits superatomic properties, with the electronic configurations being $1S^21P^61D^5$ at 1.3 Å and $1S^21P^62S^11D^4$ at 0.9 Å. Notably, superatomic behavior persists even at 1.9 Å, with an electronic configuration of $1S^21P^61D^5$. In contrast, when the radius is 2.2 Å, the MOs are not be completely delocalized within the system and distortion of the MOs becomes pronounced, indicating that the superatomic state cannot be maintained (FIG. 1(g)). These T properties become even more pronounced at a radius of 2.8 Å (FIG. 1(h)), suggesting that the $H_{13}$ system is no longer in a superatomic state when the radius exceeds 2.2 Å. Therefore, during the radius compression from 4.0 Å to 2.2 Å, the system maintains a strong electronic localization through the atom-centered MO properties. When the radius is compressed to 1.9 Å, the systems transitions into a electronic delocalization, marking the emergence of the superatomic state.

Additional studies were conducted on $M_{13}$ clusters (M = Li, Na, K) to facilitate comparison with the $H_{13}$ cluster. The results reveal that these clusters adopt an icosahedral geometry, with an electronic configuration of $1S^2 1P^6 1D^5$ (FIG. 1(i)- (k)). Notably, each 1D SAMO is half-filled with singly occupied electrons, which contributes to enhanced stability under $I_h$ symmetry. Furthermore, the MOs exhibit electron delocalized properties, further confirming that the $M_{13}$ clusters are superatomic systems.

In order to investigate the changes in the spatial distribution of electrons in $H_{13}$ during compression, we analyzed the radial distribution function (RDF) for various MOs. As shown in FIG. 2, on the basis of the evolution of curve shapes in the total RDF, the process can be divided into three regions (I, II and III). To clearly illustrate the changes in electron distributions, four representative curves from each region are displayed. In the initial compression region (region I, r ≥ 2.0 Å), the total RDF exhibits a bimodal distribution, which is also clearly reflected in the RDF curve of the 1S MO. Upon compression to r = 1.9 Å (region II, 1.2 Å < r < 2.0 Å), partial overlap between the secondary and primary peaks emerges. With further compression (from 1.9 Å to 1.3 Å), the intensity of the secondary peak in the total RDF gradually diminishes and completely vanishes at r = 1.3 Å. At r = 1.2 Å (region III, r ≤ 1.2 Å), the secondary peak disappears entirely from both the total RDF and the RDF curve of the 1S MO. The total RDF shows that a primary peak consistently appears near the radial distance. Meanwhile, a secondary peak remains around 0.5 Å, close to the Bohr radius of hydrogen (~0.53 Å), indicating a localized electron distribution around the central hydrogen atom. Therefore, in region I, the electrons are primarily localized on both the central hydrogen atom and the surrounding $H_{12}$ cage. Upon entering region II, the secondary peak begins to merge with the primary peak, suggesting that electrons initially localized on the central atom start to delocalize toward the $H_{12}$ cage. In region III, the secondary peak completely disappears and the primary peak broadens, indicating that the electrons are delocalized across the $H_{13}$ system. In addition, the RDF curves for the 1S, 1P and 1D MOs reveal that only the 1S MO holds a different number of peaks with compression, suggesting that the evolution of the total RDF is mainly dominated by the 1S MO.

Furthermore, for understanding the intrinsic mechanism of RDF peak variation, we analyzed the electron density during the compression process by selecting the maximum cross-section including the hydrogen atom at the symmetry center and the four edge hydrogen atoms (FIG. 3(a)). Specifically, to characterize the evolution of electron localization in the system, the critical electron density corresponding to the van der Waals (vdW) radius of hydrogen was calculated based on the RDF definition. The critical electron density ($\rho(r_{vdW}) = 0.0158$ e·Å$^{-3}$) was obtained by integrating over the entire hydrogen atom (see part 1.3 for computational details in Supplemental Material [33]). On the basis of this critical value (corresponding to the boundary contours), the electron density of the $H_{13}$ system was plotted (FIG. 3(a)). At a radius of 2.0 Å, the isosurfaces form isolated closed contours, indicating electron localization around individual atoms. When the radius is compressed to 1.9 Å, the emergence of overlapping isosurfaces suggests the onset of electron delocalization. This transition is consistent with the total RDF evolution shown in FIG. 2, further supporting the electron delocalization behavior under compression. Upon further compression to 1.7 Å, the overlapping regions substantially increase, indicating enhanced electron delocalization. Notably, the electron density reveals more pronounced delocalization in the central region of the system at a radius of 1.2 Å compared to 1.3 Å. This observation is further supported by the localized orbital locator (LOL) analysis (see Supplementary FIG. S2 for details [33]). These results confirm that, in the superatomic state, the electron of the central H atom is donated outward, with all 13 electrons being delocalized in the SAMOs.

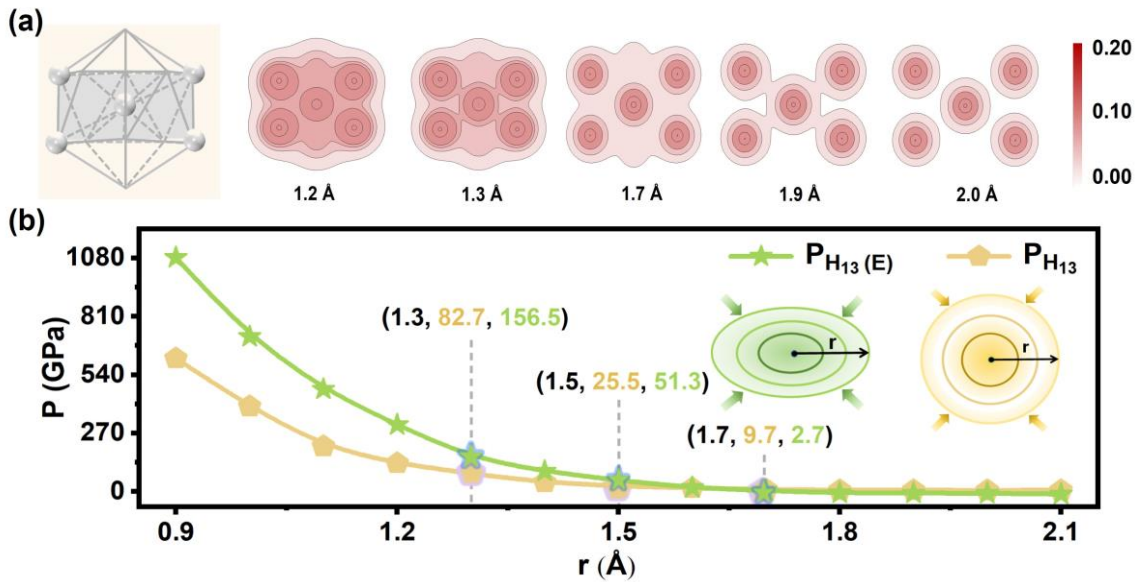

FIG. 3 Structural properties of the $H_{13}$ system during compression. (a) Cross-sectional plane selected for electron density analysis, located at the plane defined by the five atoms shown, with the gray background indicating the position of this plane. The electron density at radii of 1.2, 1.3, 1.7, 1.9, and 2.0 Å is shown on this plane. (b) Evolution of applied pressure during the compression of $H_{13}$ and ellipsoidal $H_{13}$ systems. The (E) represents the ellipsoid term.

Considering the simulation of the experimental isotropic compression process, the relationship between structural deformation and the externally applied pressure was investigated. In the pressure calculation, the external pressure (P) acting on the system is defined as $P=|dE/dV|$, where $dE$ is the total energy change and $dV$ is the total volume change during compression. Here, the system's van der Waals volume, calculated using the Bader method, is adopted. It should be noted that this approach neglects non-elastic deformation energy and differences in compression behavior between individual molecules and bulk materials, which may introduce uncertainties in pressure estimation. Nevertheless, this simplified approach provides a reasonable approximation of experimental conditions to some extent while making the calculation feasible. Therefore, the required external pressure during compression was derived using the above formula. As shown in FIG. 3(b), the required external pressure increases with the degree of compression. The pressure values required for the $H_{13}$ system were illustrated in the superatomic state at radii of 1.3, 1.5 and 1.7 Å, which are 82.7, 25.5 and 9.7 GPa, respectively. Notably, when the system radius is compressed to 1.9 Å, the required external pressure is just 6.2 GPa. These results demonstrate that the required pressure for the $H_{13}$ system in the superatomic state is two orders of magnitude lower compared to those reported in previous high-pressure experiments on metallic hydrogen. Therefore, the superatomic state may provide an avenue to reducing the external conditions for hydrogen aggregation.

All the above discussions are based on the spherical $H_{13}$ with $I_h$ symmetry, which is a relatively idealized model. Therefore, it is necessary to consider the potential impact of external inhomogeneous pressures and Jahn–Teller effect, etc. Here, these factors are briefly classified as causing structural deformation, primarily transforming ideal spherical symmetry into degenerate ellipsoidal symmetry [44-46]. In fact, such considerations are also in line with the previous understanding of changes in the electronic structure of superatoms based on ellipsoidal models [27,47]. This prompts us to consider whether the change in electronic configuration could influence the pressure required for hydrogen aggregation. In addition, the absence of a central atom in the structure may also influence the electronic configuration. Therefore, in order to gain a deeper understanding of the pressure conditions for hydrogen aggregation in different environments, spherical $H_{12}$, ellipsoidal $H_{12}$, and ellipsoidal $H_{13}$ model structures were also analyzed (see part 1.4 of Supplemental Material and FIG. S1 for details [33]), and compared with those of spherical $H_{13}$.

Through detailed analyses, we found that similar to the case of spherical $H_{13}$, the pressures required for hydrogen aggregation in the superatomic state for ellipsoidal $H_{13}$, spherical $H_{12}$, and ellipsoidal $H_{12}$ are all lower than the pressure predicted for traditional metallic hydrogen. Specifically, the SAMO shapes for these systems imply that the electronic distributions in three systems remain delocalized (see Supplementary FIG. S3 [33]), indicating their superatomic state. Based on the same estimation method as that applied to spherical $H_{13}$, as shown in FIG. 3(b), the calculation results reveal that the ellipsoidal $H_{13}$ system has a pressure of 2.1 GPa at a radial distance of 1.7 Å, which is markedly lower than the threshold required to synthesize metallic hydrogen. Similarly, the pressures for compressed spherical $H_{12}$ and ellipsoidal $H_{12}$ demonstrate the same trend (see Supplementary FIG. S4 for details [33]). These analyses suggest that hydrogen clusters can maintain their superatomic state under much lower pressure conditions than that of metallic hydrogen.

This work presents superatomic hydrogen by achieving through effective aggregation at pressures much lower than those required for recognized metallic hydrogen. High-precision multi-reference calculations reveal that, with increasing radial compression, the electrons of the $H_{13}$ system transition from localized to delocalized in MOs, providing evidence for the superatomic state. In this state, the central hydrogen atom of the $H_{13}$ system can donate electron to the $H_{12}$ cage, which embodies metallic behavior. Through theoretical calculations simulating isotropic compression in experiments, it was discovered that the pressure required to achieve the superatomic state is 6.2 GPa, two orders of magnitude lower than the generally accepted pressure of 500 GPa for metallic hydrogen. In addition, the caged $H_{12}$ system, as well as the ellipsoidal $H_{13}$ and $H_{12}$ systems with lower symmetry, were studied. These systems require higher pressures to achieve the superatomic state compared to the $H_{13}$ system, but the pressure trends remain below 500 GPa.

Undoubtedly, the reduction in required pressure also brings another scientific challenge: how to construct an effective spatial confinement system that enables the formation of a superatomic state. On the basis of previous research, such confinement conditions may favor encapsulating hydrides within clusters (such as icosahedral clusters), where the 1s electrons of hydrogen atoms collectively contribute to the total electron numbers of the clusters [48-50]. It means that the electrons of hydrogen atoms participate in the formation of the electronic structure of the superatom and ultimately achieve a maintained superatomic state. Furthermore, in the field of high-temperature superconductivity, a prominent research focus is on metal-centered hydrogen clathrate structures. Under high-pressure conditions, such configurations facilitate atomic aggregation. While the system maintains spatial translational symmetry, distinct centrosymmetry emerges at the microscopic structural level [15,18,51]. This phenomenon may open an avenue for achieving the aggregation of elemental hydrogen atoms under confinement conditions, thus presenting prospects for advancements in both fundamental principles and applied processes.

The authors wish to acknowledge Ms. R. Xu for discussion. This work was supported by the Science and Technology Development Program of Jilin Province of China (20250102014JC) and National Natural Science Foundation of China (grant numbers 11974136 and 11674123). Z. Wang also acknowledges the assistance of the High-Performance Computing Center of Jilin University and National Supercomputing Center in Shanghai.